\documentclass[3p,twocolumn]{elsarticle}

\usepackage{lineno,hyperref}
\modulolinenumbers[5]

\journal{Journal of \LaTeX\ Templates}

\begin{document}

\begin{frontmatter}

\title{Atomic structure of amorphous SiN: combining Car-Parrinello and Born-Oppenheimer first-principles molecular dynamics}

\author[ICube]{Achille Lambrecht}

\address[ICube]{Universit\'e de Strasbourg, CNRS, Laboratoire ICube, UMR 7357, F-67037 Strasbourg, France}

\author[IPCMS]{Carlo Massobrio}

\address[IPCMS]{Universit\'e de Strasbourg, CNRS, Institut de Physique et Chimie des Mat\'eriaux de Strasbourg, UMR 7504, Strasbourg F-67034, France}

\author[IPCMS]{Mauro Boero}
\author[IPCMS]{Guido Ori}

\author[ICube]{Evelyne Martin}

\ead{evelyne.martin@unistra.fr}

\begin{abstract}
First-principles molecular dynamics is employed to describe the atomic structure of amorphous SiN, 
a non-stoichiometric compound belonging to the Si$_x$N$_{y}$ family. To produce the amorphous state via the cooling of 
the liquid, both the Car-Parrinello and the Born-Oppenheimer approaches are exploited to obtain  a system featuring 
sizeable atomic mobility. At high temperatures, due to the peculiar electronic structure of SiN, exhibiting gap closing
effects, the Car-Parrinello methodology could not be followed since non-adiabatic effects involving the ionic and 
electronic degrees of freedom do occur. This shortcoming was surmounted by resorting to the  Born-Oppenheimer 
approach allowing to achieve significant ionic diffusion at $T$= 2500 K. From this highly diffusive sample, an
amorphous state at room temperature was obtained with a quenching rate of 10 K/ps.
Four different models were created, differing by their sizes and the thermal cycles. We found 
that the subnetwork of atoms N has the same environment than in the stoichiometric material Si$_3$N$_4$ since N is  
mostly threefold coordinated with Si. Si atoms can also be found coordinated to four N atoms as in Si$_3$N$_4$, but 
a substantial fraction of them forms homopolar bonds with one, two, three and even four Si. Our results are not too 
dissimilar from former models available in the litterature but they feature a higher statistical accuracy and refer 
more precisely to room temperature as the reference thermodynamical condition for the analysis of the structure in 
the amorphous state. 
\end{abstract}


\end{frontmatter}

%\linenumbers

\section{Introduction}

Amorphous silicon nitride is a dielectric material widely used in nanotechnology as an insulator\cite{BurrRecentProgressPhaseChange2016}, 
a mask for silicon etching or a cantilever of atomic force microscopes among other applications. Non-stoichiometric 
amorphous SiN films produced by chemical vapor deposition have a reduced internal stress as compared to stoichiometric 
Si$_3$N$_4$, thereby conferring them specific properties tunable with composition. As a prerequisite to a  
thermal transport study that will be performed as in recent, breakthrough investigations (see, for instance the case of
amorphous Ge$_2$Sb$_2$Te$_5$ \cite{DuongFirstprinciplesthermaltransport2021}), we have selected amorphous SiN 
(Si$_x$N$_y$ with $x=y$) as a representative non-stoichiometric silicon nitride system. This compound has remarkable thermomechanical properties \cite{FtouniThermalconductivitysilicon2015} and it is widely used in microelectronics and micromechanics manufacturing.
In this paper we construct and characterize amorphous SiN models of unprecendented quality by specifically addressing, 
from a rigourous methodological point of view, both  the production of the amorphous state and the statistical 
significance of data collected at room temperature. 

To fully appreciate the appropriateness and the legitimacy of our motivations, a few considerations are in order. 
First, it has to be underlined that first-principles molecular dynamics (FPMD) based on density functional theory (DFT) is 
the method of choice to improve upon empirical potentials \cite{DasmahapatraModelingamorphoussilicon2018} intrinsically unable 
to account for the presence of homopolar bonds at non-stoichiometric compositions. This is particularly true for covalent and/or iono-covalent systems, not easily described by analytical model potentials
based on formal charges assigned to the so-called cationic or anionic sites. However, this mere statement is not sufficient 
to ensure that a successfull handling of the interactions will be obtained at any temperature via FPMD. 

In this context, this work demonstrates that a proper treatment of SiN aimed at producing a thermal cycle cannot be achieved by relying on  the Car-Parrinello (CP) scheme since an increasing lack of adiabaticity affecting electronic and  ionic degrees of freedom, particularly severe at high temperatures, undermines this approach. 
Accordingly, the Born-Oppenheimer methodology has also been applied at the higher temperatures, making SiN an interesting 
reference case of system needing an adaptive use of FPMD strategies to be studied in a wide range of temperatures.
One could argue that the Born-Oppenheimer methodology could have been employed from the beginning instead of resorting to the Car-Parrinello approach. While this is true in principle, we found this latter method more tractable and closer in spirit to
the production of temporal trajectories based on a rigorously conserved constant of motion, fully in line with the principles of statistical mechanics. This will appear more clearly in the following of the paper.

Turning to the FPMD results already available in the literature (by Hintzsche et al \cite{HintzscheDensityfunctionaltheory2012} 
and Jarolimek et al \cite{JarolimekAtomisticmodelshydrogenated2010}), it has to be acknowledged that they have both provided
valuable information on the structure and electronic properties of amorphous SiN and, as such, they stand as fundamental 
references to account for. However, we argue that the production of the amorphous structure is not entirely convincing in 
any of these publications, 
hence calling for improvements. 
These are crucial to pursue any other study of additional properties and, in particular, thermal ones. 
In Ref. \cite{HintzscheDensityfunctionaltheory2012} a periodic model of SiN with $N$= 200 atoms is melted at $T$=4000 K 
followed by a step of 20 ps at 3000 K before undergoing temperature reduction via steps of 50 ps at  
smaller temperatures ending at $T$= 2000 K. The diffusion coefficient reported at 2500 K for SiN is close to that of a 
liquid (slightly below 10$^{-5}$ cm$^2$ s$^{-1}$) and mobility is termed as "frozen" at $T$= 2000 K.  
The question arises on whether or not, despite the vanishing diffusion at $T$=2000 K, the description  of an amorphous system carried out at such a high temperature can be taken as representative of what happens at,
say, room temperature. Indeed, it can be argued that some relaxation can still occur at the higher temperatures (in the 
quoted case, $T$= 2000 K) or, even at $T$= 300 K, when considering different periods of structural relaxation. 
This point has been largely documented, for instance, in the framework of FPMD studies on amorphous GeSe$_2$ \cite{PhysRevB.77.144207}. 
In that paper, it was shown that there are substantial changes between two sets of pair correlation functions even at room 
temperature, the first set of results being calculated at the beginning of the trajectory created via quench from the 
liquid state (the first 5 ps) and the second over the last 12 ps. 
The above rationale applies also to the FPMD models proposed by Jarolimek $et$~$al$ \cite{JarolimekAtomisticmodelshydrogenated2010} 
on hydrogenated amorphous SiN. Leaving aside the non impactful presence of hydrogen on the final 
structure \cite{HintzscheDensityfunctionaltheory2012}, the thermal history of the amorphous system features a final step  
at 300 K lasting only 0.5 ps. This results in insufficiently accurate structural properties and, again, on possible 
fingerprints of higher temperatures still affecting the network topology due to incomplete relaxation.

Given these premises and by having in mind the production of fully 
relaxed amorphous SiN system at room temperature allowing 
for a reliable calculation of thermal properties, we have planified a FPMD strategy expected to be based, in principle, on the 
Car-Parrinello \cite{CarUnifiedApproachMolecular1985} framework only. 
Our final outcome is twofold since {\it a)} we are able 
to provide robust information (on trajectories lasting 50 ps) on the structural properties of amorphous SiN at room temperature 
via a reliable thermal cycle and {\it b)} we can describe how to circumvent the inadequacies of the Car-Parrinello methodology 
encountered at high temperatures by switching to the Born-Oppenheimer approach.
The reasons underlying the use of an alternative FPMD scheme and its practical implementation have to be considered, 
by themselves, instructive achievements obtained in the context of FPMD simulations of disordered systems.
 
The paper is organized as follows. Our methodology is described in two sections, the Sec.(\ref{SecMethod1}) devoted to 
the implementation (theoretical framework, technical details of the calculations) and the Sec. \ref{SecMethod2} to a 
reminder of some conceptual issues related to the choice of the FPMD scheme. 
The temporal trajectories are detailed in Sec.\ref{SecMethod3}. 
The structural analysis of amorphous SiN is presented in Sec. \ref{SecStructure} that comprises three parts, reporting on the 
partial pair correlation functions, the coordination units and one specific representation (the Bhatia-Thornton) for the 
partial structure factors. 
Conclusive remarks are collected in Sec.  \ref{SecConclusion}.

\section{Calculations methodology I: setting up the models}
\label{SecMethod1}
\begin{figure}
\centering\includegraphics[width=8cm]{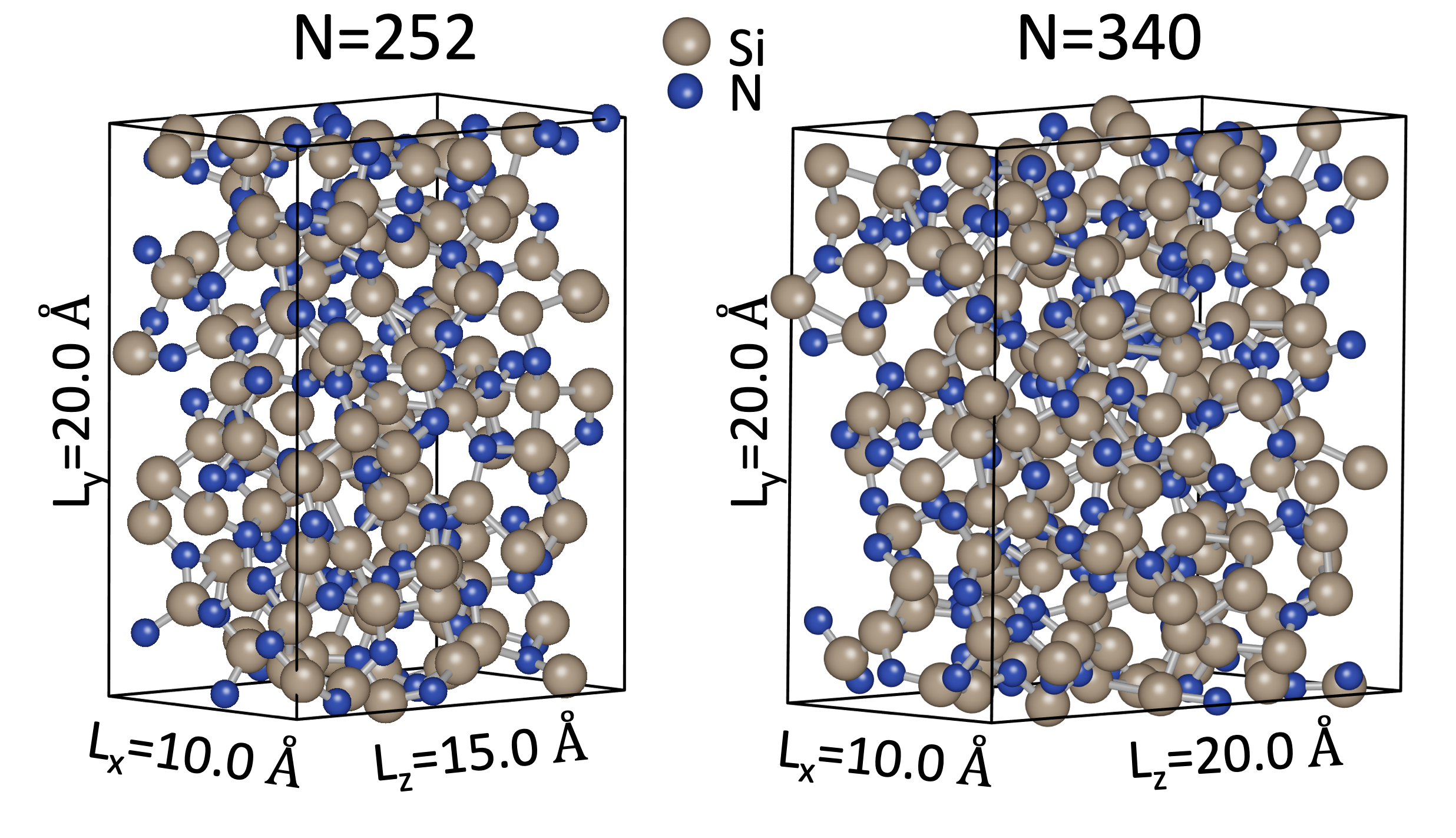}
\caption{Atomic models of  amorphous SiN containing $N$=252 and $N$=340 atoms. $L_x$, $L_y$ and $L_z$ are the dimensions of the box in the three directions.}
\label{FigModels}
\end{figure}

Periodic atomic models containing $N$=252 and $N$=340 atoms are  built at the experimental density of 2.98 g cm$^{-3}$ 
\cite{HintzscheDensityfunctionaltheory2012}. The cells are orthorombic with dimensions 10.0$\times$20.0$\times$15.0 \AA$^3$ 
and 10.0$\times$20.0$\times$20.0 \AA$^3$. 
These values are selected to ensure proper application of the approach-to-equilibrium molecular dynamics (AEMD) methodology\cite{lampin_thermal_2013,bouzid_thermal_2017,palla_interface_2019,DuongThermalconductivitytransport2019,DuongFirstprinciplesthermaltransport2021,MartinThermalconductivityamorphous2022}. 
The models employed are shown in Fig. \ref{FigModels}.  

Within DFT, one has to make a choice for the exchange-correlation part of the Kohn-Sham total energy expression. We selected 
the exchange formula proposed by Becke \cite{becke_density-functional_1988} and the correlation one of Lee, Yang and Parr \cite{LeeDevelopmentColleSalvetticorrelationenergy1988a} (BLYP). Valence-core interaction are described by norm-conserving pseudopotentials as 
prescribed by Troullier and Martins \cite{troullier_efficient_1991}. Valence electrons are represented by a plane-wave basis set 
compatible with periodic boundary conditions, with a cutoff of 50 Ry, and expanded at the $\Gamma$ point only. 
For the Car-Parrinello scheme, the mass of the fictitious electronic degrees of freedom was set to 800 a.u. and the time step to 
5 a.u. (0.12 fs) to achieve optimal conservation of the constants of motion. The ionic temperature was controlled with a 
Nos\'e-Hoover \cite{nose_molecular_1984,Noseunifiedformulationconstant1984,hoover_canonical_1985} thermostat chain \cite{MartynaNoseHooverchains1992}.  
At the beginning of our simulations, atoms are positioned randomly by making sure that unphysical configurations are avoided. We have taken advantage of a conspicuous set of disordered systems considered in recent years to select plausible initial configurations \cite{PhysRevB.86.224201, bouzid_origin_2015, doi:10.1063/1.4803115}. However,  a thermal cycle at high temperatures (temperatures at which the atoms diffuse substantially over affordable intervals of time) has to be implemented to loose memory of the initial configuration and attain a liquid phase from which to begin quenching toward an amorphous state.

\section{Calculations methodology II: choosing the appropriate FPMD scheme}
\label{SecMethod2}
 Atomic trajectories are created in the framework of  FPMD by resorting to two different and yet complementary methodologies. The first is the Car-Parrinello (CP) \cite{CarUnifiedApproachMolecular1985} method, based on a Lagrangian that depends on both ionic positions and fictitious electronic degrees of freedom (also termed in what follows ``electronic'' variables) carrying a fictitious mass. We recall that this latter set of degrees of freedom (not to be mistaken for the actual electrons) evolves in time by following adiabatically the movement of the ions, thereby greatly reducing the computational cost since no minimization of the electronic structure is needed at each time step of the ionic evolution. The search of the electronic ground state is replaced by an adjustement of the ``electronic'' variables that follow their own equations of motion with a fictitious mass (much smaller than the ionic one) so as to ensure self-consistency. Since by ``adiabatically'' one means that no energy is transferred from the ``electronic'' to the ionic degrees of freedom, this methodology stands as the optimal choice in case of systems exhibiting  a band gap at the temperatures of interest. Should this not be the case, the adiabatic conditions cannot be satisfied on arbitrarily extended time trajectories by  leading to a breakdown of the whole methodology, at least in its original formulation. When this happens, one has to resort to the Born-Oppenheimer (BO) approach (inspired by the seminal work ubiquitously referred to in any textbook of quantum mechanics \cite{BornZurQuantentheorieMolekeln1927}) that consists in a dynamical evolution of the ions under the action of forces calculated, for each time step, at the electronic ground state. 
 Unlike in the CP method, no introduction of fictitious electronic degrees of freedom is needed by allowing, in principle, the use of larger time steps for the dynamical evolution, the only dynamical frequencies to be handled being the ionic ones. It should be stressed that there are  no restrictions to the strategy of running a full FPMD-BO set of calculations at any temperature, despite the fact that the conservation of the total energy is somewhat affected by the search of the ground state electronic structure at each time step. Also, the extent of the time step in BO can affect the overall computational effort by preventing, in specific situations, any gain with respect to the CP alternative. Overall,   the comparable efficiency of the two methodologies is still contentious and do not need to be further considered here.  In this paper we employ both schemes because the second (BO) allows working at ionic temperatures not accessible to the first one (CP), this combined choice confirming the validity of both. 
We exploit the CP and BO implementations coded in the developers version 4.3 of the CPMD code \cite{cpmd}.
For a description of their conceptual framework and technical implementation see Ref. \cite{marx_hutter_2009}.

\section{Dynamical evolutions}
\label{SecMethod3}

\begin{figure*}[htb]
\centering\includegraphics[width=0.9\linewidth]{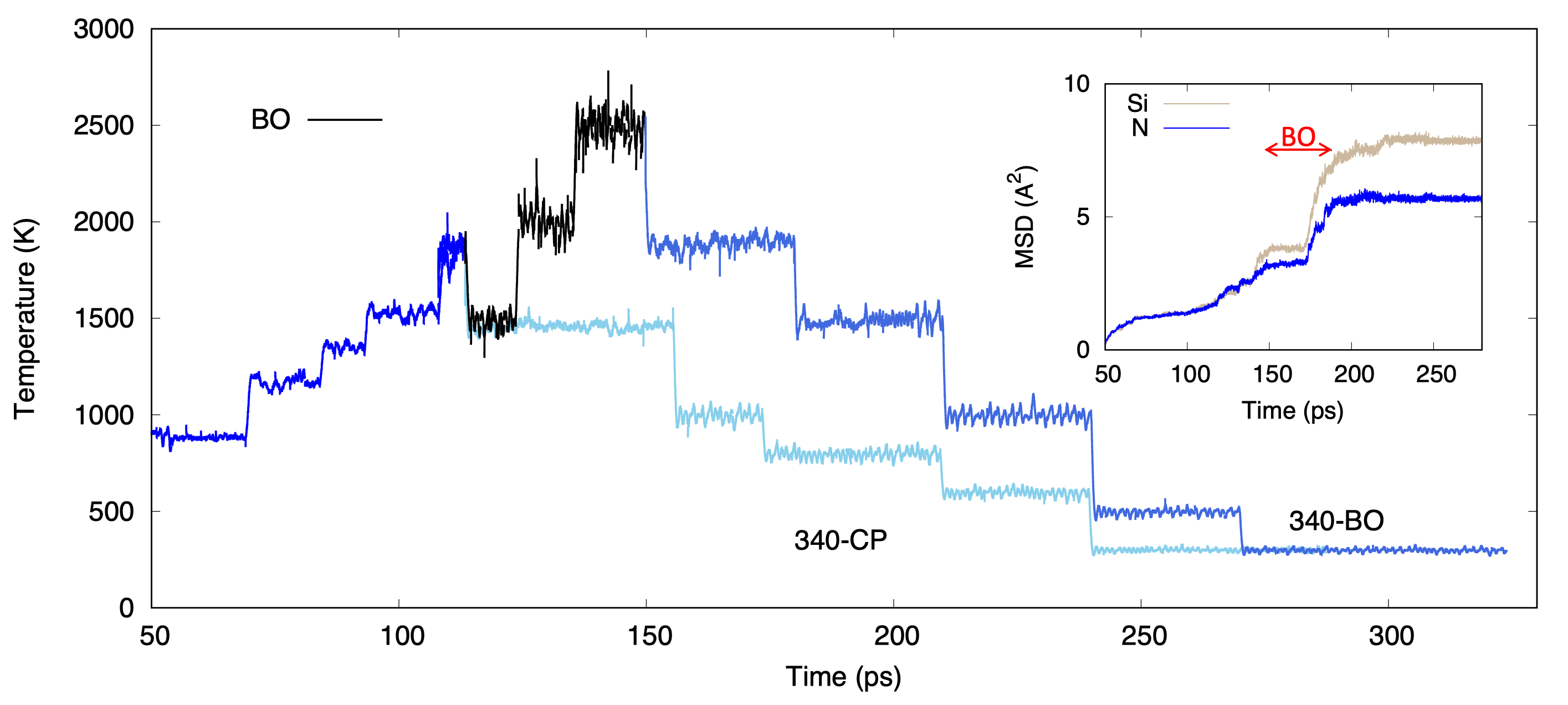}
\caption{Thermal cycles of the $N$=340 cell. The steps in blue are performed using the CP scheme, in black with the BO one. Two ramps down were performed and resulted in two amorphous models: 340-BO and 340-CP. We plot points resulting from sub-averages over 5000 atomic time units. In inset: Mean square displacement (MSD) of the Si and N atoms. The red arrow indicates the part of the thermal cycles obtained using the BO scheme. Note that the plots do not contain the
first 50 ps ($T$ lower than 900 K).}
\label{FigThC340}
\end{figure*} 

\begin{figure}[p]
\centering
\includegraphics[width=0.9\linewidth]{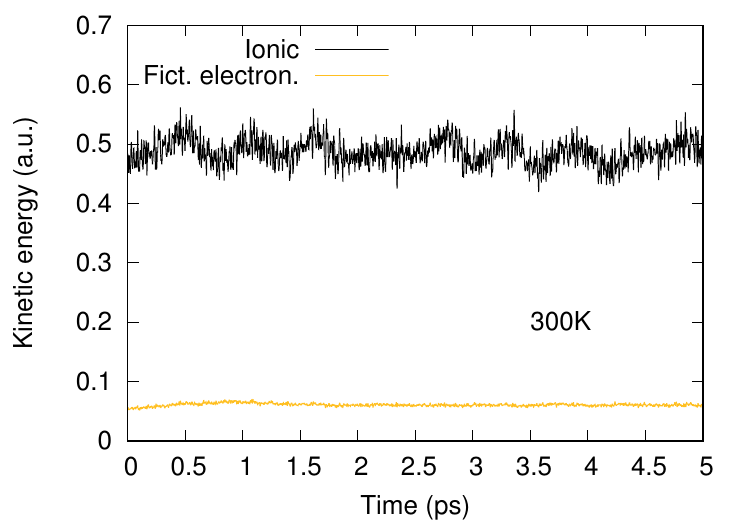}
\includegraphics[width=0.9\linewidth]{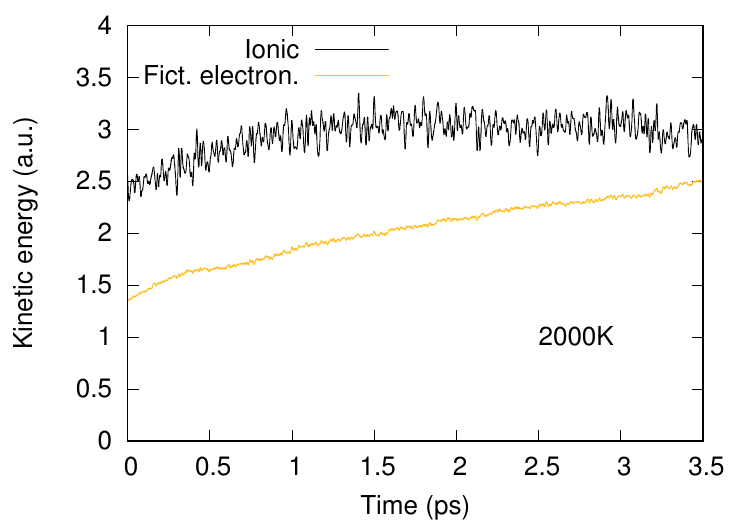}
\caption{Ionic (black curves) and fictitious electronic (gold curves) kinetic energies at $T$= 300 K (top) and 2000 K (bottom).  $N$=340.}
\label{FigKinNRJ}
\end{figure}

\begin{figure}[p]
\centering
\includegraphics[width=0.9\linewidth]{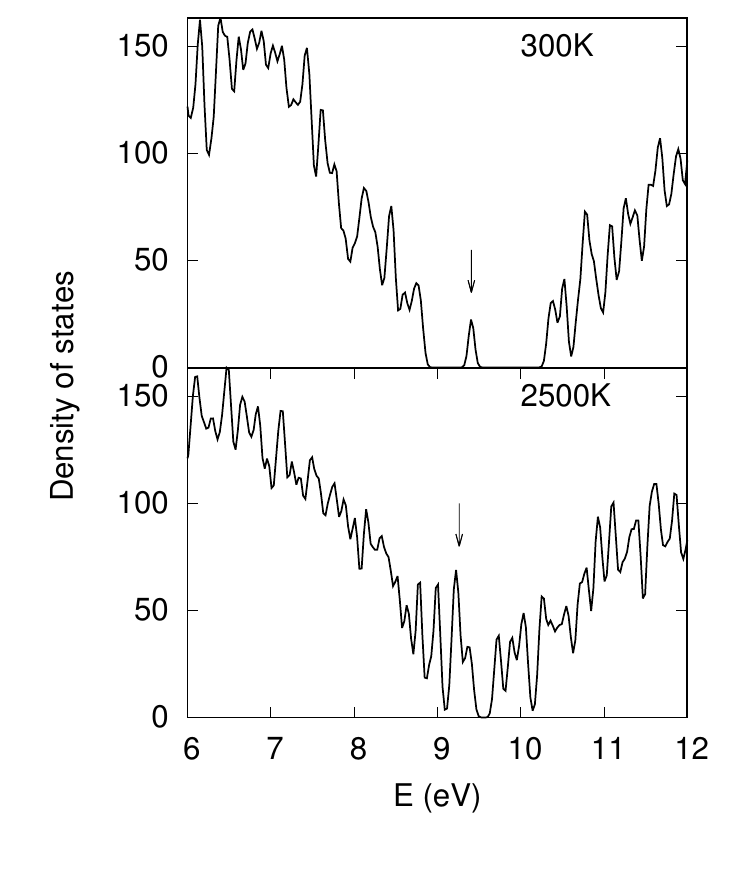}
\caption{Electronic density of states at the vicinity of the gap at 300 and 2500 K. A gaussian broadening of 50 meV has been applied. The arrows indicate the energy of the highest occupied eigenstate. The system considered  contains $N$=340 atoms.}
\label{FigeDOS}
\end{figure}

Having established that the production of an amorphous state requires the availability of a liquid structure bearing no memory 
of the initial conditions, we started increasing slowly the temperature as illustrated in Fig. \ref{FigThC340} for the system
with $N$=340 atoms. The mean square displacement of the Si and N atoms  along the trajectory for temperatures above 900 K (inset of Fig. \ref{FigThC340})
remains below 2 to 3 \AA$^2$ even when the temperature is increased up to 1600 K. At this temperature the diffusion coefficient is
$\approx 5\times$10$^{-7}$ cm$^2$ s$^{-1}$, in line with the results of Ref. \cite{HintzscheDensityfunctionaltheory2012} 
reporting $ 5\times$10$^{-5}$ cm$^2$ s$^{-1}$ at 2500 K. 
Therefore, it appears that enhanced mobility typical of a liquid state can only be obtained via a further rise of the temperature. However, at $T$= 2000 K one begins facing the intrinsic limits of validity of the CP methodology, since the kinetic energy of the  ficititious electronic degrees of freedom  tends to diverge in time (bottom part of Fig. \ref{FigKinNRJ}) and to behave non-adiabatically \cite{PastoreTheoryinitiomoleculardynamics1991} as it does, instead, at  $T$=300 K (top part of Fig. \ref{FigKinNRJ}). The observed phenomenon can be ascribed to gap closing in the electronic density of states as shown in  Fig. \ref{FigeDOS}. 

For these reasons, we turned to the Born-Oppenheimer FPMD methodology to be able to work  at higher temperatures while ensuring that the electronic structure of the system follows the self-consistent solution of the Kohn-Sham equations. 
The BO part of the full FPMD thermal cycle  is indicated in black in the temperature $vs$ time plot of Fig. \ref{FigThC340}. From the technical point of view, we found convenient to start the BO dynamics at $T$= 1500 K in order to optimize the simulation parameters before increasing the temperature to 2500 K. The most appropriate choices for the time step and the level of convergence
$\Delta E_\mathrm{tot}$  of the electronic structure at its ground state were set to 100 a. u. and 2$\times$10$^{-6}$ a. u. respectively. This leads to an acceptable compromise between the computational effort and  the conservation of the total energy, fluctuations being smaller than 5$\times$10$^{-4}$. We note that the BO framework allows a time step much larger than the CP one. However, a non negligible time is spent to reach the electronic ground state for each ionic configuration, since the conservation of the total energy was found very much dependent on the value of $\Delta E_\mathrm{tot}$ and it can rapidly worsen for 
larger values of it.
 At $T$ = 2500 K  the mean square displacements range in between $\approx$3 and $\approx$6 \AA$^2$ for  N   and in between $\approx$4 and $\approx$8 \AA$^2$ for  Si (inset of Fig. \ref{FigThC340}). Note that the higher diffusion of the Si atoms is expected since  the SiN network lies on the Si rich side of the concentration range, leading to Si atoms in homopolar bonds having higher mobility than when connecting to four N atoms. From this point on,
the quenching schedule is performed entirely in the CP framework by lowering the temperature first at the highest value compatible with this methodology ($T$ = 2000 K) and then down to $T$ = 300 K 
 for a total duration of $\approx$ 175 ps, i.e. a quench rate of 12 K/ps. The structure obtained is termed
340-BO.  A second structure (system 340-CP) is also produced by using the CP methodology and beginning to quench at  2000 K, with a longer trajectory at $T$= 1500 K (45 ps). The corresponding quench rate is equal to 10 K/ps, which falls in the acceptable range to ensure a good quality of the amorphous
phase at least for Si-based system \cite{XueEffectsquenchrates2008}.

A similar approach is applied to the simulation cell containing 252 atoms.
 The ramp up is performed in Car-Parrinello molecular dynamics up to $T$ = 2000 K, by switching to the BO one  at $T$ = 2500 K during 10 ps to boost ion diffusion. The temperature is then decreased to 300 K in 110 ps for the fast (f) quench (system 252-f) or 170 ps for the slow (s) quench (system 252-s).

\section{Structural characterization of the models}
\label{SecStructure}
\subsection{Partial pair correlation functions}
\begin{figure}[p]
\centering
\includegraphics[width=0.9\linewidth]{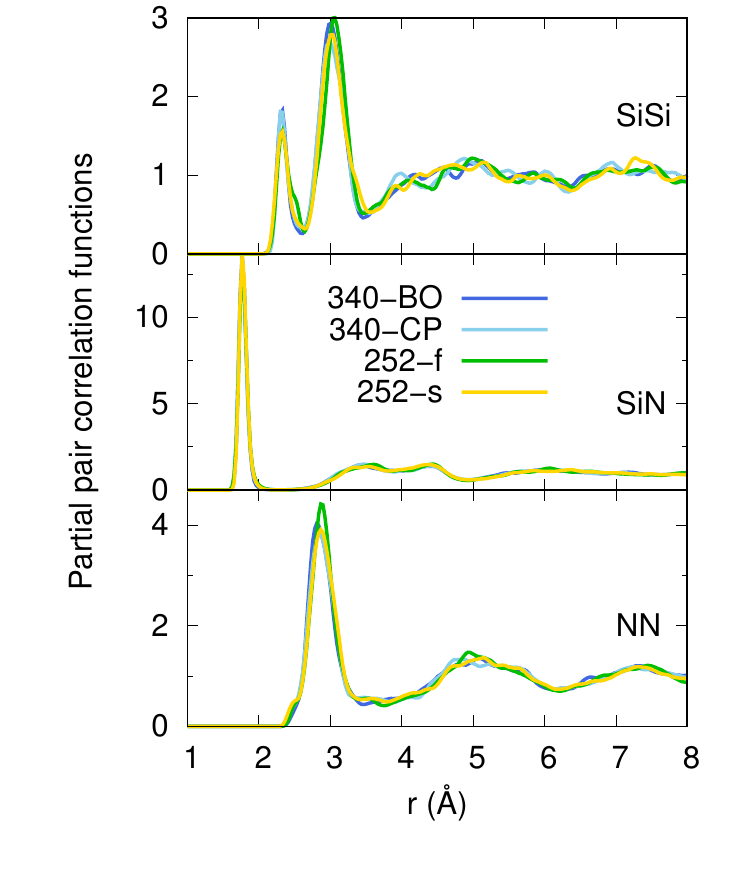}
\caption{Partial pair correlation functions of the four models of amorphous SiN obtained in the present work.}
\label{FigPCF}
\end{figure}

The partial pair correlation functions (PCFs)  $g_\mathrm{SiSi}(r)$,  $g_\mathrm{NN}(r)$ and  $g_\mathrm{SiN}(r)$ calculated for the four amorphous systems
 are shown in Fig. \ref{FigPCF}. 
Overall, the four sets of   PCFs share the same features, as confirmed by  the positions of the peaks and the coordination numbers $n_{\alpha\beta}$ reported respectively in Table \ref{TabDist} and  \ref{TableCN}.
\begin{table}[htb]
\centering
\caption{Nearest neighbors distances in \AA\ between Si-N, Si-Si and N-N atoms in amorphous silicon nitride.}
\label{TabDist}
\begin{tabular}{c || c | c  |c | c | c}
& Si-N&\multicolumn{2}{c |}{Si-Si}&\multicolumn{2}{c}{N-N}\\
&1$^\mathrm{st}$&1$^\mathrm{st}$&2$^\mathrm{nd}$ &1$^\mathrm{st}$&2$^\mathrm{nd}$\\
\hline
\hline
\multicolumn{6}{l}{Si$_3$N$_4$ exp.}\\
\hline
Ref. \cite{AiyamaXraydiffractionstudy1979} &1.75&$-$&3.00 &$-$&3.00\\
Ref. \cite{MisawaStructurecharacterizationCVD1979}&1.73&$-$&3.01 &$-$&3.00\\
\hline
\multicolumn{6}{l}{SiN present work ($\pm$ 0.02 \AA)}\\
\hline
252-f &1.76&2.33&2.99 &$-$&2.86\\
252-s&1.76&2.33&3.04 &$-$&2.86\\
340-CP &1.76&2.30&3.00 &$-$&2.87\\
340-BO &1.76&2.33&3.00 &$-$&2.81\\
\hline
\multicolumn{6}{l}{SiN previous results}\\
\hline
Ref. \cite{HintzscheDensityfunctionaltheory2012}  &1.75&2.38&3.02 &$-$&2.89\\
Ref. \cite{JarolimekAtomisticmodelshydrogenated2010}&1.745&2.352&3.030 &$-$&2.854\\
\hline
\hline
\end{tabular}
\end{table}

\begin{table}[htb]
\centering
\caption{Coordination numbers $n_{\alpha\beta}$ obtained by integrating up to the first minima of the pair correlation functions g$_{\alpha\beta}(r)$}
\label{TableCN}
\begin{tabular}{c || c   |c | c }
& $n_\mathrm{SiN}$&$n_\mathrm{SiSi}$&$n_\mathrm{NN}$\\
\hline
\hline
\multicolumn{4}{l}{Present work }\\
\hline
252-f &3.02&1.19 &0.0\\
252-s&3.02&1.14 &0.0\\
340-CP &3.02&1.12 &0.0\\
340-BO &3.02&1.12 &0.0\\
\hline
\multicolumn{4}{l}{Chemically ordered network}\\
\hline
 &3&1 &0\\
 \hline
\multicolumn{4}{l}{Previous results}\\
\hline
Ref. \cite{HintzscheDensityfunctionaltheory2012}  &2.99&1.01&0.0\\
Ref. \cite{JarolimekAtomisticmodelshydrogenated2010}&2.90&0.88&0.0\\
\hline
\hline
\end{tabular}
\end{table}

 The Si-N distance is equal to 1.76 $\pm$ 0.02 \AA, in agreement with the values reported in Refs. \cite{HintzscheDensityfunctionaltheory2012}  and \cite{JarolimekAtomisticmodelshydrogenated2010}. The value corresponds to the distance in the stoichiometric materials as measured in Refs. \cite{AiyamaXraydiffractionstudy1979} and
\cite{MisawaStructurecharacterizationCVD1979}. The first peak of the $g_\mathrm{SiN}(r)$ has very close  amplitudes and widths in our four models, while  $g_\mathrm{SiSi}(r)$ has a double peak, centered at 2.32 $\pm$ 0.02 \AA~and 3.00$\pm$ 0.02 \AA~as average values.  The coordination numbers corresponding to the integration of these peaks up to the first minimum are reported in Table \ref{TableCN}.  There is no first Si-Si peak to compare with in the stoichiometric Si$_3$N$_4$ as observed in  Refs.  \cite{HintzscheDensityfunctionaltheory2012}  and \cite{JarolimekAtomisticmodelshydrogenated2010}. Note that   the available Si-Si distance of amorphous SiN found in those studies is slightly larger than in our case (2.38  \AA\  and 2.35 \AA\ vs 2.32  \AA\  in average). The second peak of $g_\mathrm{SiSi}(r)$ is located at the distance defining the second shell of neighbors in Si$_3$N$_4$, $\sim$ 3.0 \AA\ for all models.  $g_\mathrm{NN}(r)$ has a single peak, but we label it ``2$^\mathrm{nd}$ dist" in Table \ref{TabDist} since it is not a signature of  nearest neighbour positions ($\approx$ 1.6 \AA) observed in overstoichiometric silicon nitride  \cite{HintzscheDensityfunctionaltheory2012}. In our model, the average N-N distance is equal to 2.85 \AA\, this values being slightly lower (2.81 \AA) for the model 340-BO.

\subsection{Coordinations}

 The coordinations numbers $n_{\alpha\beta}$ compiled in Table \ref{TableCN} are similar for the four models. They are close to the values obtained within the chemically ordered network model (CON). The notion of CON in disordered materials deserves some more explanations, since it is based, for each concentration, on the existence of bonds between atoms of the same or different chemical nature, so as to maximize the number of heteropolar ones. When considering  Si$_x$N$_{1-x}$ systems, stoichiometry is ensured when $x$= 3/7 and, in this case (Si$_3$N$_4$), a perfect chemical order occurs when each Si atom has four  N neighbors and each N atom has three Si neighbors, with neither Si$-$Si nor N$-$N homopolar bonds. Moving out of stoichiometry at the $x$=0.5 concentration, the  chemically ordered network (CON) model implies that  only Si$-$N and Si$-$Si bonds are allowed for  $x >$ 3/7 while the opposite would be true (Si$-$N and N$-$N bonds allowed) for $x <$ 3/7. 
 
 The coordination numbers are close to those obtained in Refs. \cite{HintzscheDensityfunctionaltheory2012} and \cite{JarolimekAtomisticmodelshydrogenated2010}, although the numbers are slightly smaller in the presence of hydrogen (Ref. \cite{JarolimekAtomisticmodelshydrogenated2010}). 
 In order to elucidate the environment of Si and N atoms, one has to focus on the coordinations in more details via the individual ${n}_\alpha(l)$ structural units, where an atom of species $\alpha$ (say, Si or N) is $l$--fold coordinated to other atoms, not all of them necessarily belonging to the same species.

Within this notation, 1N2Si${n}_\mathrm{Si}(3)$ refers to Si atoms  connected to 1 N atom and 2 Si atoms while 4N${n}_\mathrm{Si}(4)$ corresponds to Si atoms connected to 4 N atoms.  Therefore
 ${n}_\alpha(l)$ is  the number of atoms of a given species that are coordinated to $l$ atoms, the average being calculated over the entire trajectory.  By calculating  ${n}_\alpha(l)$ one has access to more information than the one conveyed by the coordination numbers, since  the entire range   of the interatomic distances (accessible to the periodic box) becomes available. This allows describing  the coordination environment of species $\alpha$ at any  $r$. All neighbors of any kind within a sphere of radius $r$ are included in the counting of the interactions that concur to the calculations of ${n}_\alpha(l)$.
 
\begin{figure}[p]
\centering
\includegraphics[width=0.9\linewidth]{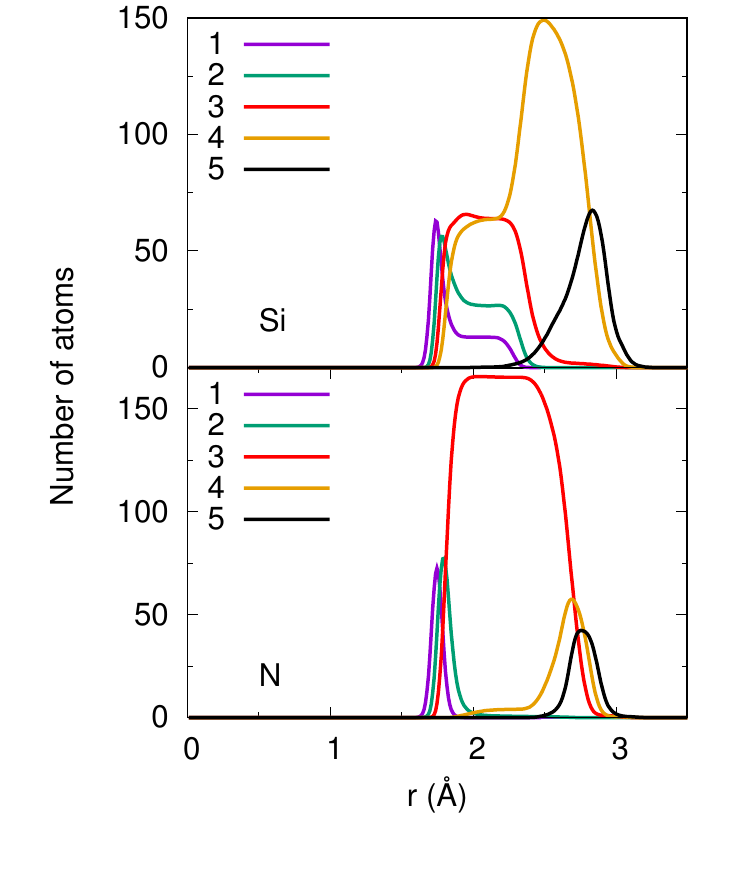}
\caption{Coordination units ${n}_\mathrm{Si}(l)$ and ${n}_\mathrm{N}(l)$ atoms (bottom) for $l$= 1,2,3,4,5. System 340-BO at 300 K.}
\label{FigCNtot}
\end{figure}

\begin{figure}[htb]
\centering
\includegraphics[width=0.9\linewidth]{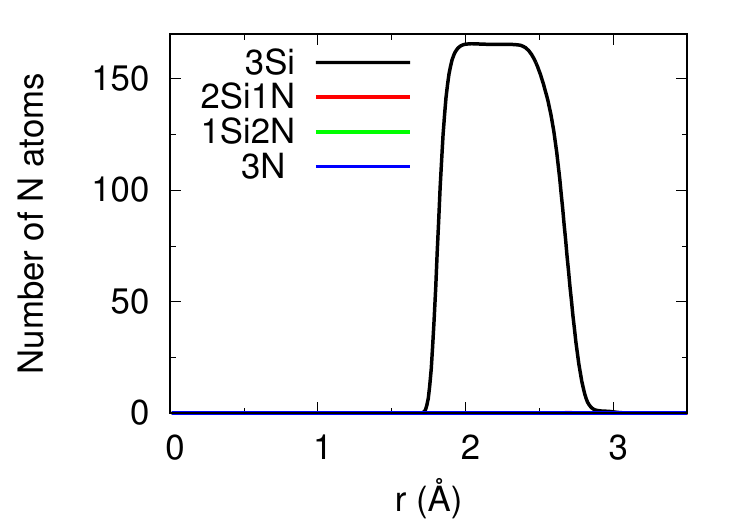}
\caption{Decomposition of ${n}_\mathrm{N}(3)$ by accounting for the chemical identity of the $l$ neighboring atoms. We have  3Si${n}_\mathrm{N}(3)$ (black line), 2SiN${n}_\mathrm{N}(3)$ (red line), Si2N${n}_\mathrm{N}(3)$ (green line) and 3N${n}_\mathrm{N}(3)$ (blue line). The system considered is 340-BO at 300 K.}
\label{FigCN3}
\end{figure}

 For the system 340-BO, Fig. \ref{FigCNtot} gives ${n}_\alpha(l)$ for $l$=1, 2, 3, 4 and 5 and
 $\alpha$= Si or N  within a range of relevant interatomic distances. The results are averaged over the final step of the thermal cycle at 300 K ($\approx$ 50 ps). In the case of N, the overwhelming majoity of atoms are threefold coordinated over the range comprised between
 2~\AA~and 3~\AA.   At short $r$, ${n}_\mathrm{N}(1)$ and ${n}_\mathrm{N}(2)$ are also non negligible, as it occurs for ${n}_\mathrm{N}(4)$ and ${n}_\mathrm{N}(5)$ when $r$  approaches the first minimum of $g_\mathrm{NN}(r)$. These results are complemented by those shown in Fig. \ref{FigCN3}, where it appears clearly that the structural motif accounting for ${n}_{\mathrm{N}}(3)$ is 3Si${n}_{\mathrm{N}}(3)$. The atoms of species N behave largely as  in stoichiometric Si$_3$N$_4$ , i. e. they form three bonds with silicon atoms. Note that the plateau in ${n}_\mathrm{N}$ is consistent with the behavior of $g_\mathrm{SiN}(r)$, featuring values very close to zero in between 2 and 3~\AA.

\begin{figure}[p]
\centering
\includegraphics[width=0.9\linewidth]{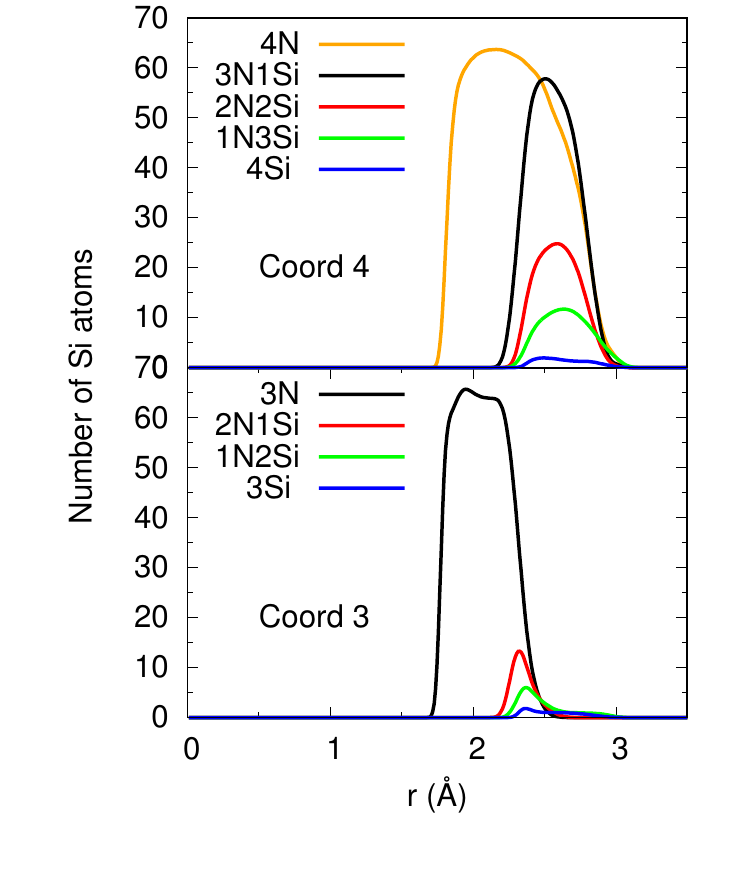}
\caption{Top part: decomposition of ${n}_\mathrm{Si}(4)$ by accounting for the chemical identity of the $l$ neighboring atoms. 4N${n}_\mathrm{Si}(4)$ (orange line), 3N1Si${n}_\mathrm{Si}(4)$ (black line),  2N2Si${n}_\mathrm{Si}(4)$ (red), 1N3Si${n}_\mathrm{Si}(4)$ (green) and 4Si${n}_\mathrm{Si}(4)$ Si (blue). 
Bottom part: decomposition of ${n}_\mathrm{Si}(3)$ by accounting for the chemical identity of the $l$ neighboring atoms. 3N${n}_\mathrm{Si}(3)$ (black line), 2N1Si${n}_\mathrm{Si}(3)$ (red line),  1N2Si${n}_\mathrm{Si}(3)$ (green line),  and 3Si${n}_\mathrm{Si}(3)$ (blue line)
The system considered is 340-BO at 300 K.}
\label{FigCSi34}
\end{figure}

In the case of Si, ${n}_\mathrm{Si}(4)$ is the dominant coordination unit (Fig. \ref{FigCNtot}). The pattern recorded for ${n}_\mathrm{Si}(1)$ and ${n}_\mathrm{Si}(2)$  corresponds to the shortest bonds formed with the nearest neighbours, while ${n}_\mathrm{Si}(5)$ is related to Si neighbors of the second shell  as shown by the peak at 3 \AA\ on $g_\mathrm{SiSi}(r)$. 
Being responsible for the largest contributions of Si, the coordination units ${n}_\mathrm{Si}(3)$ and ${n}_\mathrm{Si}(4)$ are worth some close attention by referring to  Fig. \ref{FigCSi34}. 
By looking at  ${n}_\mathrm{Si}(4)$ (top part of Fig. \ref{FigCSi34}), it appears that there is a large amount of Si atoms coordinated to four N, as in the stoichiometric material. The shape of
${n}_\mathrm{Si}(4)$ is remindful of  ${n}_\mathrm{N}(3)$ (Fig. \ref{FigCN3}). In addition, Si is also found in 3N1Si${n}_\mathrm{Si}(4)$ configurations, these occurring at  larger distances since involving the Si neighbors that account for first peak of $g_\mathrm{SiSi}(r)$ (2.33 \AA). In order to understand how these features can coexist, let us look at the various ${n}_\mathrm{Si}(3)$ units (bottom part of Fig. \ref{FigCSi34}). The predominant structural unit is 3N${n}_\mathrm{Si}(3)$. This threefold Si coordination, standing out  in Fig. \ref{FigCSi34} (bottom part), is in reality part of a fourfold Si coordination made of 3 N atoms at typical Si-N distances (1.76 \AA) and one Si-Si bond at a larger distance (2.33 \AA). This is exactly what it was found via the observation of
the top part of Fig. \ref{FigCSi34}. It remains true that in addition to this 3N1Si${n}_\mathrm{Si}(4)$ unit, Si atoms can also be found in a sizeable proportion of 4N${n}_\mathrm{Si}(4)$ units as in the stoichiometric material. Also, other units like 2N2Si${n}_\mathrm{Si}(4)$, 1N3Si${n}_\mathrm{Si}(4)$ and,
in vary small proportions, 4Si${n}_\mathrm{Si}(4)$ can also be encountered.

\begin{figure}[htb]
\centering
\includegraphics[width=\linewidth]{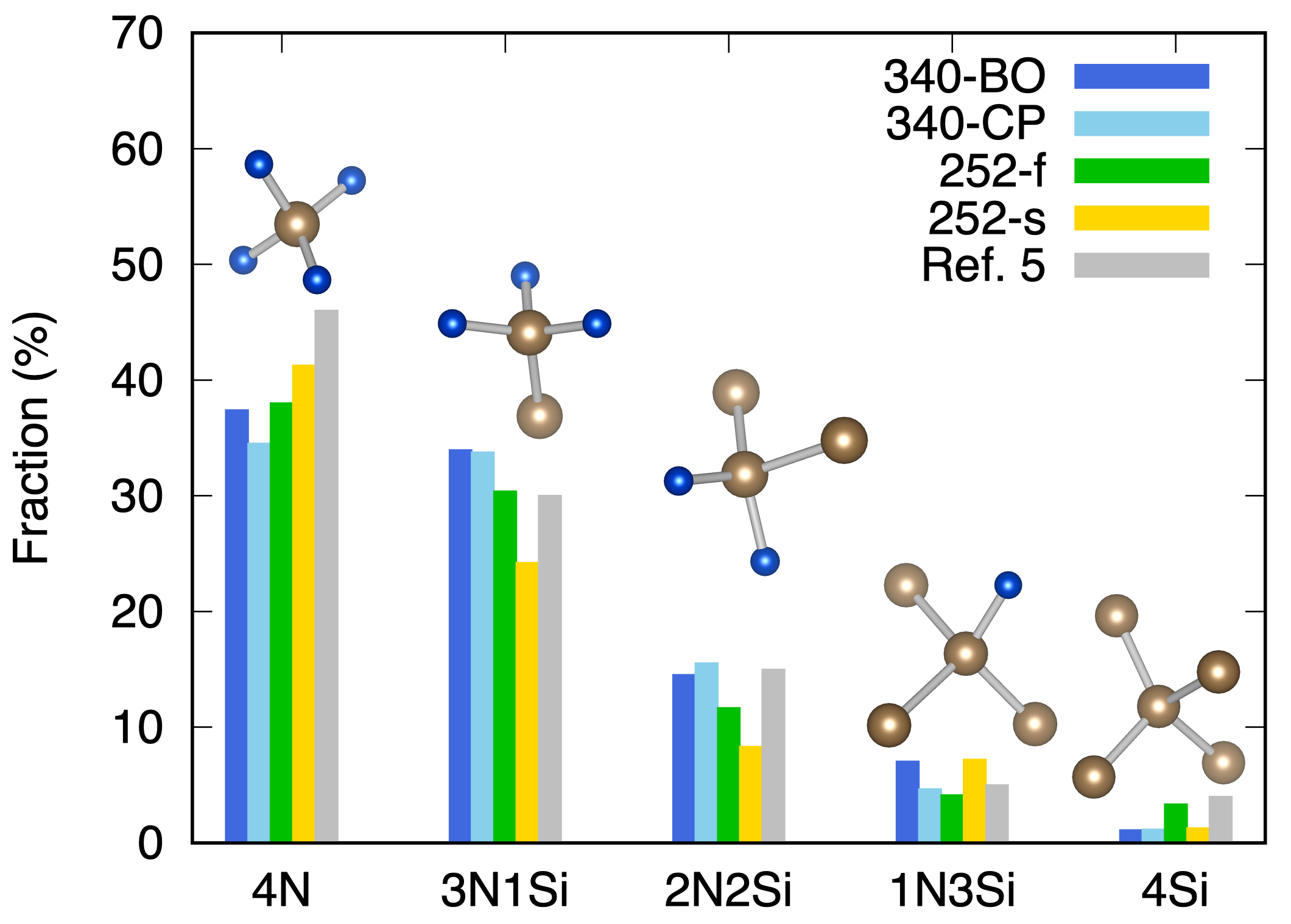}
\caption{Fractions of the various environnments of fourfold coordinated Si atoms in the four models and comparison with the results obtained in Ref. \cite{HintzscheDensityfunctionaltheory2012}. The Si are pictured as brown balls and the N atoms as blue ones.}
\label{FigFracSi4}
\end{figure}

The distribution of the various occurrencies for the different coordination units ${n}_\mathrm{Si}(4)$  is given in Fig. \ref{FigFracSi4}. The calculation was performed for the four systems. The results are comparable and do not depend either on the size or on the details of the thermal cycles. Between 35 and 41 \% of Si atoms are coordinated to 4 N atoms as in the stoichiometric material. The remaining Si atoms form homopolar bonds with the excess Si atoms,  the number of neighbors of the same kind including up to four connections. While this feature was also in Ref. \cite{HintzscheDensityfunctionaltheory2012})  there is no compelling reason to interpret it as a signature of phase separation.  This would imply a sharper split of   ${n}_\mathrm{Si}(4)$ in two or three sub-categories (4N${n}_\mathrm{Si}(4)$ against 4Si${n}_\mathrm{Si}(4)$ or 1N3Si${n}_\mathrm{Si}(4)$) as opposed to the four distinct classes observed in which both N (heteropolar bonding) and Si (homopolar bonding) coexist as neighbors of Si.  The presence of Si and N as neigbors of Si proves that the different units are cross-linked thereby preventing the formation of two topologically distinct networks.

To summarize on the coordination units of our amorphous system, our network is chemically ordered (as shown in Tab. \ref{TableCN} by comparison with the chemically ordered network model (CON)) and it features N atoms mostly coordinated with four Si atoms, while Si atoms accomodate to a large extent in two structural units, by forming bonds with four N atoms or with three N atoms and one Si atoms, located at larger distances.

\subsection{Bhatia-Thornton partial structure factors}
\begin{figure}[htb]
\centering
\includegraphics[width=\linewidth]{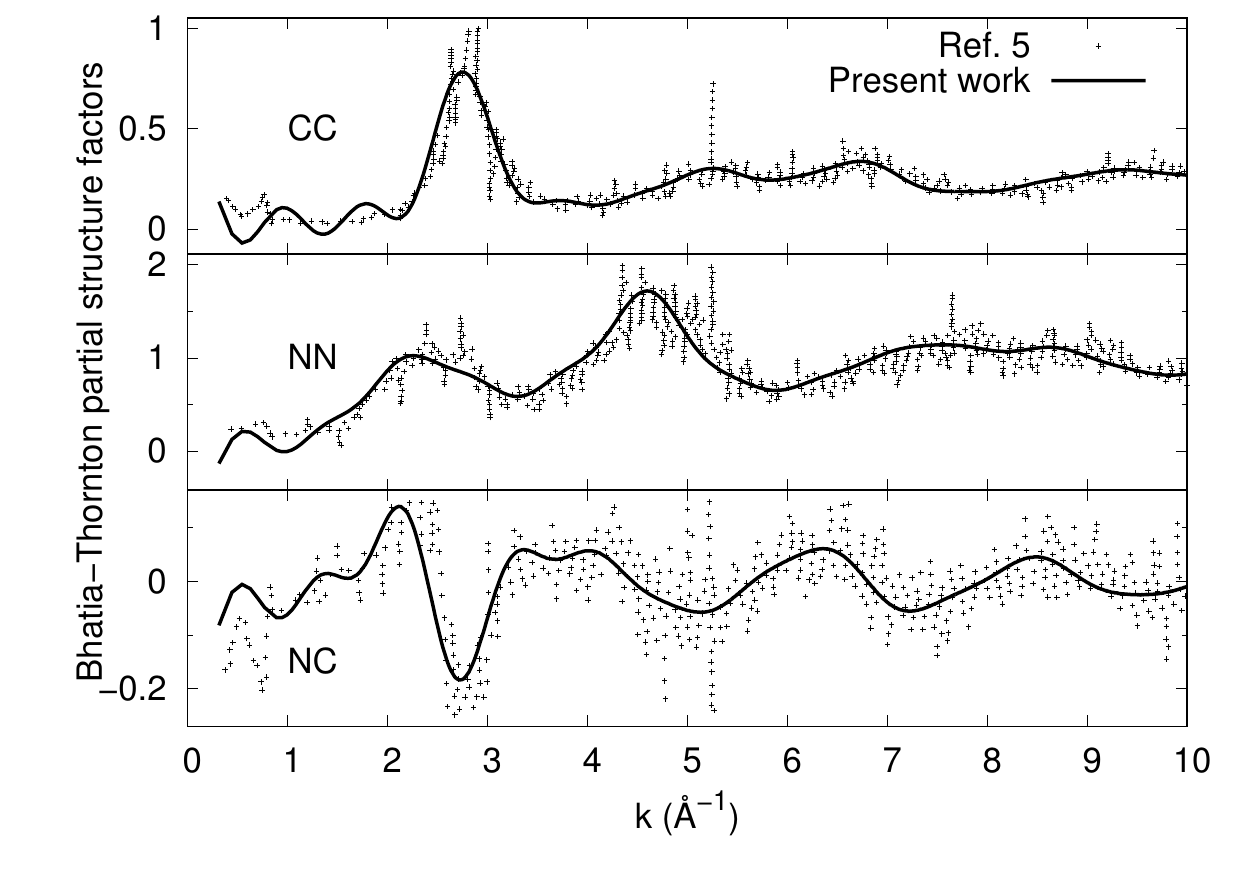}
\caption{Bhatia-Thornton partial structure factors averaged over the four models of amorphous SiN models (340-BO, 340-CP, 252-f and 252-s) (line) and results from the calculation of Ref. \cite{JarolimekAtomisticmodelshydrogenated2010}.}
\label{FigBT}
\end{figure}

In addition to being complementary to pair correlation functions, partial structure factors  provide information on the existence of extended order beyond nearest neighbors (intermediate range order). This  manifests itself through the appearance of a peak at low $k$ values (typically around 1 \AA$^{-1}$)  located at the left of the main peak.
Several amorphous do exhibit such a feature, named FSPD (first sharp diffraction peak), in the total and in some of the partial structure factors, calling for a wealth of interpretations of its atomic scale origin  \cite{PhysRevLett.67.711,doi:10.1063/1.1365108}.
  In addition to a description in terms of atomic species, involving direct and cross correlations,
expressed via the set of the so-called Faber-Ziman  structure factors \cite{WasedaStructureNoncrystallineMaterials1980}, one can focus on the notions of number-number, number-concentration and concentration-concentration partial structure factors, built in the Bhatia and Thornton formalism \cite{Salmonstructuremoltenglassy1992}. 
In Fig. \ref{FigBT}, we show the partial structure factors $S_\mathrm{NN}(k)$ (number-number),
$S_\mathrm{NC}(k)$ (number-concentration) and $S_\mathrm{CC}(k)$ (concentration-concentration).
These can be obtained by linear combinations of the Faber-Ziman  structure factors \cite{WasedaStructureNoncrystallineMaterials1980}
as follows, where $\alpha$= Si and $\beta$= N

\begin {eqnarray}\label{SNN}
S_\mathrm{NN}(k)= c_{{\alpha}}c_{{\alpha}}S_{\alpha\alpha}(k)\nonumber\\ + 2 c_{{\alpha}}c_{{\beta}}S_{\alpha\beta}(k) 
+ c_{{\beta}}c_{{\beta}}S_{\beta\beta}(k)
\end{eqnarray}

\begin {eqnarray}\label{SNC}
S_\mathrm{NC}(k)= c_{{\alpha}}c_{{\beta}}\{c_{{\alpha}}
 (S_{\alpha\alpha}(k)\nonumber\\ -S_{\alpha\beta}(k)) -c_{{\beta}}(S_{\beta\beta}(k) -S_{\alpha\beta}(k))\}.
\end{eqnarray}

\begin {eqnarray}\label{SCC}
S_\mathrm{CC}(k)= c_{{\alpha}}c_{{\beta}}\{1 + c_{{\alpha}}c_{{\beta
}}[
 (S_{\alpha\alpha}(k)\nonumber\\-S_{\alpha\beta}(k))+(S_{\beta\beta}(k)-S_{\alpha\beta}(k))]\}.
\end{eqnarray}

While $S_\mathrm{NN}(k)$ is indicative of a global effect due to all species, its significance being quite similar to that of the total neutron structure factor  commonly measured for amorphous systems, $S_\mathrm{NC}(k)$ and $S_\mathrm{NC}(k)$ reflects the different weigths of the direct and cross structure factors, with peaks appearing at given $k$ when the system is sensitive to the distinct chemical nature of its components. Depending on the relative sign of these contributions, $S_\mathrm{NC}(k)$ and $S_\mathrm{NC}(k)$ can be either positive or negative, their trend following roughly opposite patterns. 
Fig. \ref{FigBT} refers to results obtained by Fourier transformation of the pair correlation functions to obtain first the Faber-Ziman partial structure factors and then converted into the Bathia-Thornton expressions via
Eq. \ref {SNN},  Eq. \ref {SNC} and Eq. \ref {SCC}. Averages are taken over the four models produced (340-BO, 340-CP, 252-f, 252-s).
Overall, it appears that our data are much less noisy than those obtained in Ref. \cite{JarolimekAtomisticmodelshydrogenated2010} due to the combined effect of much longer trajectories and the use of a Fourier transformation, smoothing the data obtained directly in reciprocal space.  Based on what shown in Fig. \ref{FigBT}, we found no particular marks  of  intermediate range order in any  partial structure factor, this meaning that for distances beyond nearest neighbors there are no correlation effects related to number and/or concentration effects as expressed in Eqs. \ref{SNN}, \ref{SNC} and \ref{SCC}. 

\section{Conclusion}
\label{SecConclusion}
The structural properties of amorphous SiN$_x$, $x=1$ have been obtained by exploiting two first-principles molecular dynamics (FPMD) methodologies on periodic cells of 252 and 340 atoms. In this respect, this work has a twofold importance, since an accurate structural analysis is obtained via a careful choice of the best suited theoretical approach, providing an instructive example of combined use of FPMD methods for disordered systems. 
Four models have been constructed by applying different thermal cycles aimed at producing an amorphous configuration at room temperature.  Our calculations improve upon previous FPMD results due to unprecedented lengths for the temporal trajectories composing the thermal cycle  and the attainment of a network configurationally arrested precisely at room temperature. In order to loose memory of the initial configuration and produce a melted disordered state, we increased the temperature up to $T$=2500 K, thereby facing intrinsic drawbacks due to the  Car-Parrinello methodology (lack of adiabaticity for the combined electronic/ionic evolution scheme). For this reason, we resorted to the Born-Oppenheimer approach that allowed us to produce trajectories sampling adequately the phase space so as to obtain significant atomic mobility. Then, the systems are carefully quenched to room temperature with a quench rate of 10 K/ps. The analysis of the structure shows that  the N atoms have the same environment than in the stoichiometric material Si$_3$N$_4$, i.e. they form three bonds with Si atoms.  Si atoms are found in a variety of configurations, $\sim$ 30 \% of them being coordinated to four N as in Si$_3$N$_4$. When found in the fourfold coordination, Si exhibits also homopolar connections characterized by bond distance larger than in the heteropolar case. 

Overall, our results on the atomic structure of amorphous SiN do not differ substantially from those obtained
 in Ref.~\cite{JarolimekAtomisticmodelshydrogenated2010} and Ref.~\cite{HintzscheDensityfunctionaltheory2012}. 
However, Ref.~\cite{HintzscheDensityfunctionaltheory2012} suffered from a lack of data on the amorphous structure at room temperature, being based on the assumption that one can describe such topology by ending the simulations at $T$= 2000 K.
Concerning the results of Ref.~\cite{JarolimekAtomisticmodelshydrogenated2010}, they were obtained by exploiting a very short relaxation trajectory at room temperature, thereby preventing the system for any residual structural relaxation. Based on these considerations, the overall agreement between the present set of results, based on much longer temporal evolutions, and those already available is by itself an interesting outcome. In any event, this should not be taken as an indication that statistical accuracy is unimportant for this specific system.
We do prefer to point out the superior validity of our approach that is based on an unprecedented
sampling of the phase space at different temperatures and allows fully exploiting the versatility of first-principles molecular dynamics methods. 
For these reasons, they are well suited to pave the way to investigations devoted to the thermal behavior of this disordered system.

\section*{Conflicts of interest} 
There are no conflicts of interest to declare.

\section*{Acknowledgements} 

This work was funded by the French ANR via the project n. ANR-17-CE09-0039-02 ``SIRENA" and by ICube via the project ``Model\_Thermiq\_PCMem" (grant A. L.). We acknowledge PRACE for awarding us access to Joliot-Curie at GENCI@CEA, France and  GENCI (Grand Equipement National de Calcul Intensif) (Grant No. A0xx0905071, A0xx0910296). We also acknowledge the High Performance Computing Center of the University of Strasbourg for supporting this work by providing scientific support and access to computing resources. Part of the computing resources were funded by the Equipex Equip@Meso project (Programme Investissements d'Avenir) and the CPER Alsacalcul/Big Data.

\section*{Data availability}
The raw/processed data required to reproduce these findings are available from the corresponding author on reasonable request.

%\section*{References}

\bibliography{full-biblio}

\begin{thebibliography}{10}
\expandafter\ifx\csname url\endcsname\relax
  \def\url#1{\texttt{#1}}\fi
\expandafter\ifx\csname urlprefix\endcsname\relax\def\urlprefix{URL }\fi
\expandafter\ifx\csname href\endcsname\relax
  \def\href#1#2{#2} \def\path#1{#1}\fi

\bibitem{BurrRecentProgressPhaseChange2016}
G.~W. Burr, M.~J. Brightsky, A.~Sebastian, H.-Y. Cheng, J.-Y. Wu, S.~Kim, N.~E.
  Sosa, N.~Papandreou, H.-L. Lung, H.~Pozidis, E.~Eleftheriou, C.~H. Lam,
  Recent {{Progress}} in {{Phase-Change Memory Technology}}, IEEE Journal on
  Emerging and Selected Topics in Circuits and Systems 6~(2) (2016) 146--162.
\newblock \href {http://dx.doi.org/10.1109/JETCAS.2016.2547718}
  {\path{doi:10.1109/JETCAS.2016.2547718}}.

\bibitem{DuongFirstprinciplesthermaltransport2021}
T.-Q. Duong, A.~Bouzid, C.~Massobrio, G.~Ori, M.~Boero, E.~Martin,
  First-principles thermal transport in amorphous {{Ge$_2$Sb$_2$Te$_5$}} at the
  nanoscale, RSC Advances 11~(18) (2021) 10747--10752.
\newblock \href {http://dx.doi.org/10.1039/D0RA10408F}
  {\path{doi:10.1039/D0RA10408F}}.

\bibitem{FtouniThermalconductivitysilicon2015}
H.~Ftouni, C.~Blanc, D.~Tainoff, A.~D. Fefferman, M.~Defoort, K.~J. Lulla,
  J.~Richard, E.~Collin, O.~Bourgeois, Thermal conductivity of silicon nitride
  membranes is not sensitive to stress, Physical Review B 92~(12) (2015)
  125439.
\newblock \href {http://dx.doi.org/10.1103/PhysRevB.92.125439}
  {\path{doi:10.1103/PhysRevB.92.125439}}.

\bibitem{DasmahapatraModelingamorphoussilicon2018}
A.~Dasmahapatra, P.~Kroll, Modeling amorphous silicon nitride: {{A}}
  comparative study of empirical potentials, Computational Materials Science
  148 (2018) 165--175.
\newblock \href {http://dx.doi.org/10.1016/j.commatsci.2017.12.008}
  {\path{doi:10.1016/j.commatsci.2017.12.008}}.

\bibitem{HintzscheDensityfunctionaltheory2012}
L.~E. Hintzsche, C.~M. Fang, T.~Watts, M.~Marsman, G.~Jordan, M.~W. P.~E.
  Lamers, A.~W. Weeber, G.~Kresse, Density functional theory study of the
  structural and electronic properties of amorphous silicon nitrides:
  {{Si}}$_3${{N}}$_{4-x}$:{{H}}, Physical Review B 86~(23) (2012) 235204.
\newblock \href {http://dx.doi.org/10.1103/PhysRevB.86.235204}
  {\path{doi:10.1103/PhysRevB.86.235204}}.

\bibitem{JarolimekAtomisticmodelshydrogenated2010}
K.~Jarolimek, R.~A. {de Groot}, G.~A. {de Wijs}, M.~Zeman, Atomistic models of
  hydrogenated amorphous silicon nitride from first principles, Physical Review
  B 82~(20) (2010) 205201.
\newblock \href {http://dx.doi.org/10.1103/PhysRevB.82.205201}
  {\path{doi:10.1103/PhysRevB.82.205201}}.

\bibitem{PhysRevB.77.144207}
C.~Massobrio, A.~Pasquarello,
  \href{https://link.aps.org/doi/10.1103/PhysRevB.77.144207}{Short and
  intermediate range order in amorphous {G}e{S}e$_2$}, Phys. Rev. B 77 (2008)
  144207.
\newblock \href {http://dx.doi.org/10.1103/PhysRevB.77.144207}
  {\path{doi:10.1103/PhysRevB.77.144207}}.
\newline\urlprefix\url{https://link.aps.org/doi/10.1103/PhysRevB.77.144207}

\bibitem{CarUnifiedApproachMolecular1985}
R.~Car, M.~Parrinello, Unified {{Approach}} for {{Molecular Dynamics}} and
  {{Density-Functional Theory}}, Physical Review Letters 55~(22) (1985)
  2471--2474.
\newblock \href {http://dx.doi.org/10.1103/PhysRevLett.55.2471}
  {\path{doi:10.1103/PhysRevLett.55.2471}}.

\bibitem{lampin_thermal_2013}
E.~Lampin, P.~L. Palla, P.-A. Francioso, F.~Cleri, Thermal conductivity from
  approach-to-equilibrium molecular dynamics, Journal of Applied Physics
  114~(3) (2013) 033525.
\newblock \href {http://dx.doi.org/10.1063/1.4815945}
  {\path{doi:10.1063/1.4815945}}.

\bibitem{bouzid_thermal_2017}
A.~Bouzid, H.~Zaoui, P.~L. Palla, G.~Ori, M.~Boero, C.~Massobrio, F.~Cleri,
  E.~Lampin, Thermal conductivity of glassy {{GeTe$_4$}} by first-principles
  molecular dynamics, Physical Chemistry Chemical Physics 19~(15) (2017)
  9729--9732.
\newblock \href {http://dx.doi.org/10.1039/C7CP01063J}
  {\path{doi:10.1039/C7CP01063J}}.

\bibitem{palla_interface_2019}
P.~L. Palla, S.~Zampa, E.~Martin, F.~Cleri, Interface thermal behavior in
  nanomaterials by thermal grating relaxation, International Journal of Heat
  and Mass Transfer 131 (2019) 932--943.
\newblock \href {http://dx.doi.org/10.1016/j.ijheatmasstransfer.2018.11.064}
  {\path{doi:10.1016/j.ijheatmasstransfer.2018.11.064}}.

\bibitem{DuongThermalconductivitytransport2019}
T.-Q. Duong, C.~Massobrio, G.~Ori, M.~Boero, E.~Martin, Thermal conductivity
  and transport modes in glassy {{GeTe}}$_4$ by first-principles molecular
  dynamics, Physical Review Materials 3~(10) (2019) 105401.
\newblock \href {http://dx.doi.org/10.1103/PhysRevMaterials.3.105401}
  {\path{doi:10.1103/PhysRevMaterials.3.105401}}.

\bibitem{MartinThermalconductivityamorphous2022}
E.~Martin, G.~Ori, T.-Q. Duong, M.~Boero, C.~Massobrio, Thermal conductivity of
  amorphous {{SiO$_2$}} by first-principles molecular dynamics, Journal of
  Non-Crystalline Solids 581 (2022) 121434.
\newblock \href {http://dx.doi.org/10.1016/j.jnoncrysol.2022.121434}
  {\path{doi:10.1016/j.jnoncrysol.2022.121434}}.

\bibitem{becke_density-functional_1988}
A.~D. Becke, Density-functional exchange-energy approximation with correct
  asymptotic behavior, Physical Review A 38~(6) (1988) 3098--3100.
\newblock \href {http://dx.doi.org/10.1103/PhysRevA.38.3098}
  {\path{doi:10.1103/PhysRevA.38.3098}}.

\bibitem{LeeDevelopmentColleSalvetticorrelationenergy1988a}
C.~Lee, W.~Yang, R.~G. Parr, Development of the {{Colle-Salvetti}}
  correlation-energy formula into a functional of the electron density,
  Physical Review B 37~(2) (1988) 785--789.
\newblock \href {http://dx.doi.org/10.1103/PhysRevB.37.785}
  {\path{doi:10.1103/PhysRevB.37.785}}.

\bibitem{troullier_efficient_1991}
N.~Troullier, J.~L. Martins, Efficient pseudopotentials for plane-wave
  calculations, Physical Review B 43~(3) (1991) 1993--2006.
\newblock \href {http://dx.doi.org/10.1103/PhysRevB.43.1993}
  {\path{doi:10.1103/PhysRevB.43.1993}}.

\bibitem{nose_molecular_1984}
S.~Nos{\'e}, A molecular dynamics method for simulations in the canonical
  ensemble, Molecular Physics 52~(2) (1984) 255--268.
\newblock \href {http://dx.doi.org/10.1080/00268978400101201}
  {\path{doi:10.1080/00268978400101201}}.

\bibitem{Noseunifiedformulationconstant1984}
S.~Nos{\'e}, A unified formulation of the constant temperature molecular
  dynamics methods, The Journal of Chemical Physics 81~(1) (1984) 511--519.
\newblock \href {http://dx.doi.org/10.1063/1.447334}
  {\path{doi:10.1063/1.447334}}.

\bibitem{hoover_canonical_1985}
W.~G. Hoover, Canonical dynamics: {{Equilibrium}} phase-space distributions,
  Physical Review A 31~(3) (1985) 1695--1697.
\newblock \href {http://dx.doi.org/10.1103/PhysRevA.31.1695}
  {\path{doi:10.1103/PhysRevA.31.1695}}.

\bibitem{MartynaNoseHooverchains1992}
G.~J. Martyna, M.~L. Klein, M.~Tuckerman, Nos\'e\textendash{{Hoover}} chains:
  {{The}} canonical ensemble via continuous dynamics, The Journal of Chemical
  Physics 97~(4) (1992) 2635--2643.
\newblock \href {http://dx.doi.org/10.1063/1.463940}
  {\path{doi:10.1063/1.463940}}.

\bibitem{PhysRevB.86.224201}
S.~Le~Roux, A.~Bouzid, M.~Boero, C.~Massobrio,
  \href{https://link.aps.org/doi/10.1103/PhysRevB.86.224201}{Structural
  properties of glassy {Ge}$_{2}${Se}$_{3}$ from first-principles molecular
  dynamics}, Phys. Rev. B 86 (2012) 224201.
\newblock \href {http://dx.doi.org/10.1103/PhysRevB.86.224201}
  {\path{doi:10.1103/PhysRevB.86.224201}}.
\newline\urlprefix\url{https://link.aps.org/doi/10.1103/PhysRevB.86.224201}

\bibitem{bouzid_origin_2015}
A.~Bouzid, S.~Le~Roux, G.~Ori, M.~Boero, C.~Massobrio, Origin of structural
  analogies and differences between the atomic structures of {GeSe$_4$} and
  {{GeS$_4$}} glasses: {{A}} first principles study, The Journal of Chemical
  Physics 143~(3) (2015) 034504.
\newblock \href {http://dx.doi.org/10.1063/1.4926830}
  {\path{doi:10.1063/1.4926830}}.

\bibitem{doi:10.1063/1.4803115}
S.~Le~Roux, A.~Bouzid, M.~Boero, C.~Massobrio,
  \href{https://doi.org/10.1063/1.4803115}{The structure of liquid {GeSe}
  revisited: A first principles molecular dynamics study}, The Journal of
  Chemical Physics 138~(17) (2013) 174505.
\newblock \href {http://arxiv.org/abs/https://doi.org/10.1063/1.4803115}
  {\path{arXiv:https://doi.org/10.1063/1.4803115}}, \href
  {http://dx.doi.org/10.1063/1.4803115} {\path{doi:10.1063/1.4803115}}.
\newline\urlprefix\url{https://doi.org/10.1063/1.4803115}

\bibitem{BornZurQuantentheorieMolekeln1927}
M.~Born, R.~Oppenheimer, Zur {{Quantentheorie}} der {{Molekeln}}, Annalen der
  Physik 389~(20) (1927) 457--484.
\newblock \href {http://dx.doi.org/10.1002/andp.19273892002}
  {\path{doi:10.1002/andp.19273892002}}.

\bibitem{cpmd}
{Jointly by IBM Corporation and by Max Planck Institute, Stuttgart},
  \href{http://www.cpmd.org}{C{P}{M}{D} code} (2021).
\newline\urlprefix\url{http://www.cpmd.org}

\bibitem{marx_hutter_2009}
D.~Marx, J.~Hutter, Ab Initio Molecular Dynamics: Basic Theory and Advanced
  Methods, Cambridge University Press, 2009.
\newblock \href {http://dx.doi.org/10.1017/CBO9780511609633}
  {\path{doi:10.1017/CBO9780511609633}}.

\bibitem{PastoreTheoryinitiomoleculardynamics1991}
G.~Pastore, E.~Smargiassi, F.~Buda, Theory of ab initio molecular-dynamics
  calculations, Physical Review A 44~(10) (1991) 6334--6347.
\newblock \href {http://dx.doi.org/10.1103/PhysRevA.44.6334}
  {\path{doi:10.1103/PhysRevA.44.6334}}.

\bibitem{XueEffectsquenchrates2008}
K.~Xue, L.-S. Niu, H.-J. Shi, Effects of quench rates on the short- and
  medium-range orders of amorphous silicon carbide: {{A}} molecular-dynamics
  study, Journal of Applied Physics 104~(5) (2008) 053518.
\newblock \href {http://dx.doi.org/10.1063/1.2974095}
  {\path{doi:10.1063/1.2974095}}.

\bibitem{AiyamaXraydiffractionstudy1979}
T.~Aiyama, T.~Fukunaga, K.~Niihara, T.~Hirai, K.~Suzuki, An {{X-ray}}
  diffraction study of the amorphous structure of chemically vapor-deposited
  silicon nitride, Journal of Non-Crystalline Solids 33~(2) (1979) 131--139.
\newblock \href {http://dx.doi.org/10.1016/0022-3093(79)90043-7}
  {\path{doi:10.1016/0022-3093(79)90043-7}}.

\bibitem{MisawaStructurecharacterizationCVD1979}
M.~Misawa, T.~Fukunaga, K.~Niihara, T.~Hirai, K.~Suzuki, Structure
  characterization of {{CVD}} amorphous {{Si$_3$N$_4$}} by pulsed neutron total
  scattering, Journal of Non-Crystalline Solids 34~(3) (1979) 313--321.
\newblock \href {http://dx.doi.org/10.1016/0022-3093(79)90018-8}
  {\path{doi:10.1016/0022-3093(79)90018-8}}.

\bibitem{PhysRevLett.67.711}
S.~R. Elliott,
  \href{https://link.aps.org/doi/10.1103/PhysRevLett.67.711}{Origin of the
  first sharp diffraction peak in the structure factor of covalent glasses},
  Phys. Rev. Lett. 67 (1991) 711--714.
\newblock \href {http://dx.doi.org/10.1103/PhysRevLett.67.711}
  {\path{doi:10.1103/PhysRevLett.67.711}}.
\newline\urlprefix\url{https://link.aps.org/doi/10.1103/PhysRevLett.67.711}

\bibitem{doi:10.1063/1.1365108}
C.~Massobrio, A.~Pasquarello, \href{https://doi.org/10.1063/1.1365108}{Origin
  of the first sharp diffraction peak in the structure factor of disordered
  network-forming systems: Layers or voids?}, The Journal of Chemical Physics
  114~(18) (2001) 7976--7979.
\newblock \href {http://arxiv.org/abs/https://doi.org/10.1063/1.1365108}
  {\path{arXiv:https://doi.org/10.1063/1.1365108}}, \href
  {http://dx.doi.org/10.1063/1.1365108} {\path{doi:10.1063/1.1365108}}.
\newline\urlprefix\url{https://doi.org/10.1063/1.1365108}

\bibitem{WasedaStructureNoncrystallineMaterials1980}
Y.~Waseda, The {{Structure}} of {{Non-crystalline Materials}}: {{Liquids}} and
  {{Amorphous Solids}}, {McGraw-Hill International Book Company}, 1980.

\bibitem{Salmonstructuremoltenglassy1992}
P.~S. Salmon, The structure of molten and glassy 2:1 binary systems: An
  approach using the {{Bhatia}}\textemdash{{Thornton}} formalism, Proceedings
  of the Royal Society of London. Series A: Mathematical and Physical Sciences
  437~(1901) (1992) 591--606.
\newblock \href {http://dx.doi.org/10.1098/rspa.1992.0081}
  {\path{doi:10.1098/rspa.1992.0081}}.

\end{thebibliography}

\end{document}